# Improved approach to Fowler-Nordheim plot analysis

[Change of title from: "Use of a new type of intercept correction factor to improve Fowler-Nordheim plot analysis"]

Richard G. Forbes[a]

*Advanced Technology Institute & Department of Electronic Engineering, Faculty of Engineering and Physical Sciences, University of Surrey, Guildford, Surrey GU2 7XH, UK*

Andreas Fischer[b], Marwan S. Mousa

*Department of Physics, Mu'tah University, Al-Karak 61710, Jordan*

---

[a] Author to whom correspondence should be addressed; permanent alias for electronic mail: r.forbes@trinity.cantab.net

[b] Present address: Institut für Physik, Technische Universität Chemnitz, Germany.

**[Abstract]**

This article introduces an improved approach to Fowler-Nordheim (FN) plot analysis, based on a new type of intercept correction factor. This factor is more cleanly defined than the factor previously used. General enabling theory is given that applies to any type of FN plot of data that can be fitted using a FN-type equation. Practical use is limited to emission situations where slope correction factors can be reliably predicted. By making a series of well-defined assumptions and approximations, it is shown how the general formulas reduce to provide an improved theory of orthodox FN-plot data analysis. This applies to situations where the circuit current is fully controlled by the emitter characteristics, and tunneling can be treated as taking place through a Schottky-Nordheim (SN) barrier. For orthodox emission, good working formulas make numerical evaluation of the slope correction factor and the new intercept correction factor quick and straightforward. A numerical illustration, using simulated emission data, shows how to use this improved approach to derive values for parameters in the full FN-type equation for the SN barrier. Good self-consistency is demonstrated. The general enabling formulas also pave the way for research aimed at developing analogous data-analysis procedures for non-orthodox emission situations.

## I. Introduction

Fowler-Nordheim (FN) plots[1] are often used to interpret experimental data relating to cold field electron emission (CFE). This article proposes an improvement to the tangent method[2] of analyzing FN plots. The new approach is intended to replace the existing theory[2-5] of the tangent method, and to act as general enabling theory for FN plot analysis of data described empirically by a FN-type equation. An improved approach is clearly needed because most existing methods of FN plot analysis implicitly assume that the field emitting regions are planar surfaces, but this is not a valid approximation for many modern emitters.

A stimulus for this article has been recent work[6] on CFE theory for large area field emitters (LAFEs). This suggested that many LAFE papers contain a defective FN-type equation that omits necessary correction factors and over-predicts the macroscopic (i.e., LAFE-average) emission current density by a large factor, possibly by as much as $10^9$ in some cases. Methods of extracting values for FN-type-equation pre-exponentials from experimental current-voltage ($i$-$V$) data are thus of interest. The two main approaches are "spot-value" methods, as used in Ref. 6, and methods that interpret the intercept of a line fitted to a FN plot. "Simulate and compare" approaches are also possible. This article is about interpreting the slopes and intercepts of FN plots.

Many ways exist of presenting CFE data as FN plots, since any one of several "independent" variables (notably, voltage, barrier field, scaled barrier field, and macroscopic field), and any one of several "dependent" variables (notably, current, local current density, and macroscopic current density) can be used. To provide generality, a "universal" theoretical formulation is used here, in which $X$ denotes any suitable independent variable, and $Y$ any suitable dependent variable.

Physically, FN-type equations describe CFE from a metal conduction band, when the emitter tip is "not too sharp" (radius greater than of order 10 nm). However, FN-type equations are also used as empirical fitting equations for all types of field emission data, notwithstanding that interpretation of extracted parameters is sometimes problematic.

When the universal variables $X$ and $Y$ are used, alternative general forms for the technically complete[6,7] "universal" FN-type equation are:

$$Y = CX^2 \exp[-G_F] \equiv CX^2 \exp[-\nu_F S^*/X]. \tag{1}$$

The subscript "F", here and elsewhere, indicates that the parameter refers to a barrier of zero-field height equal to the local work function $\phi$. For this barrier, $G_F$ is the barrier strength (formerly called the JWKB exponent), and $\nu_F$ ("nu$_F$") is a correction factor related to the barrier's mathematical form. The precise mathematical forms of $S^*$ and $C$ depend on which independent and dependent variables are being used, and (for $C$) on what physical/modelling assumptions are being made. In so-called "FN coordinates", Eq. (1) becomes a function $L(X^{-1})$ where $L \equiv \ln\{Y/X^2\}$: the related FN plot then graphs $L$ against $X^{-1}$. Thus:

$$L(X^{-1}) \equiv \ln\{Y/X^2\} = \ln\{C\} - \nu_F S^* \cdot X^{-1}. \tag{2}$$

There is a significant difference in thinking between the new and old approaches. Older approaches focussed directly on physical interpretation of the FN plot slope and intercept (for example, on the intercept as a parameter related to emission area). The new approach separates interpretation into two parts: (1) the "extraction problem" (e.g., how to extract a value for the parameter $C$ from the measured FN-plot intercept); and (2) the "interpretation problem" (e.g., how to physically interpret this extracted $C$-value). This article is about the separation process and the extraction problem. It aims to make deriving reliable characterization parameters from experiments easier, once necessary correction factors have been found (which must be done separately).

Due to the peculiar mathematics of FN-type equations when written in FN coordinates, the FN-plot intercept is not $\ln\{C\}$ (except in elementary theory–which is physically unrealistic), and the extraction problem is far more complicated and unexplored than is generally recognized. Extracting meaningful parameter values from FN plots requires knowledge or reasonable estimation of slope and intercept correction factors for the emitter in question. The usual assumptions made in analyzing FN-plot slopes and intercepts disregard many corrections. Some corrections are obviously small, but some are not, and some have never been fully investigated. For example, a hidden assumption often made is that the emitting areas at the emitter tips can be treated as effectively planar. For many modern emitters this is unlikely to be true. Slope and intercept correction factors will be different for planar and curved emitters, but published literature contains no systematic studies of how emitter shape affects factor values.

Making the above separation allows the extraction problem to be studied in its own right. To implement this separation, a new type of intercept correction factor is defined (below) that relates only to the extraction process. This paper sets out relevant "enabling" theory in an abstract general form that applies to all FN plot types, and is general enough to deal formally with all corrections known to us.

By considering slope and intercept correction factors separately, it is then shown how (by making a defined series of assumptions and approximations) this general theory can be reduced to mathematically explicit theory applicable to the so-called[8] "orthodox emission situation". This explicit theory constitutes an improved ("more complete") form of orthodox FN-plot analysis—which is the form of analysis that applies to planar metal emitters.

The present article demonstrates—by analyzing simulated data—that this improved orthodox data-analysis theory is self-consistent, in that it extracts with good accuracy the parameter values input to the simulations. It is then argued that this approach could also be used for non-orthodox situations, provided reliable values could be found for the relevant slope and intercept correction factors. This would normally require separate specific calculations (for the chosen emitter model) of parameters defined generally in the present article.

The article structure is as follows. Section II deals with some necessary background theoretical issues; Sections III and IV discuss the definitions of slope and intercept correction factors; Section V provides a simulation-based illustration of extraction procedures; and Section VI provides discussion and a summary. For logarithms, standard curly-bracket notation is used whereby ln{$x$} means: "write the value of the physical quantity $x$ in specified units, and take the natural logarithm of its numerical value when $x$ is given in these units".

## II. BACKGROUND THEORY

### A. The auxiliary parameter $c_X$

The independent variable in formal theoretical derivations of FN-type equations is a characteristic value $F_C$ of the local barrier field. This relates to the universal variable $X$ by the auxiliary equation

$$F_C = c_X X, \qquad (3)$$

where $c_X$ is the relevant auxiliary parameter. For example, if $X$ is the macroscopic field $F_M$, then $c_X$ is the characteristic macroscopic field enhancement factor $\gamma_C$. Hence, the barrier strength $G_F$ can be written in the alternative forms

$$G_F = v_F G_F^{ET} = v_F b\phi^{3/2}/F_C = v_F b\phi^{3/2}/c_X X = v_F S^*/X, \qquad (4)$$

where $b$ is the Second FN Constant[9], $G_F^{ET}$ [$= b\phi^{3/2}/F_C$] is the barrier strength for an exact triangular (ET) barrier of height $\phi$, and $S^*$ is a positive quantity defined by

$$S^* \equiv b\phi^{3/2}/c_X. \qquad (5)$$

### B. Orthodox emission

The concept of *orthodox emission* relates to field-emitter behavior in its whole environment, including the vacuum system and the electrical measurement circuit. The full definition of orthodox emission[8] aims to provide statements of a "paradigm" body of basic CFE theory and of the ideal conditions under which it can be used to predict emitter behavior and interpret measured data. Key requirements are that the emission current must be "fully emitter controlled" (i.e., controlled solely by emission at the emitter/vacuum interface), and that emission can be treated as obeying the full FN-type equation based on tunneling through a Schottky-Nordheim (SN) barrier.

Modern emission situations may or may not be effectively orthodox. If the emission situation is not orthodox, then orthodox and elementary data-analysis methods may generate spurious results. Measured and simulated emission data can now be tested for lack of orthodoxy[8].

### C. Scaled barrier field $f$ and the parameter $\eta$

In orthodox emission theory, it is often convenient to use dimensionless "scaled" parameters[6,8,10]. The scaled barrier field $f$ (for an SN barrier of zero-field height $\phi$) is

$$f \equiv F/F_R = c_S^2 \phi^{-2} F \approx 1.439\,964\ (\phi/\text{eV})^{-2}\ (F/\text{V nm}^{-1}), \tag{6}$$

where $c_S$ is the Schottky constant[9] and the "reference field" $F_R\ [=c_S^{-2}\phi^2]$ is the field needed to reduce to zero an SN barrier of zero-field height $\phi$. The parameter $\eta$ is

$$\eta \equiv bc_S^2 \phi^{-1/2} \approx 9.836\,238\ (\text{eV}/\phi)^{1/2}. \tag{7}$$

When $\phi = 4.50$ eV, then $F_R \approx 14.06$ V/nm, and $\eta \approx 4.637$.

Merits of using $f$ and $\eta$ are that (a) the parameters are dimensionless, (b) only a single field-like variable is involved, (c) barrier field is proportional to $f$, and (d) in orthodox emission, $f$ can also be treated as a scaled voltage[11], or other relevant scaled independent variable.

## III. THE SLOPE CORRECTION FACTOR

### A. Basic theory

Let $S(X^{-1})$ be the theoretically expected slope of a FN plot of $L(X^{-1})$. For CFE, $S(X^{-1})$ is a negative quantity that varies slightly or significantly with $X^{-1}$. The *slope correction factor* $\sigma_{YX}(X^{-1})$ is a positive quantity defined by

$$\sigma_{YX}(X^{-1}) \equiv -S(X^{-1})/S^*. \tag{8}$$

The subscript "YX" is added as a reminder that, in this universal formulation, the value of $\sigma_{YX}(X^{-1})$ depends on the choices made for $X$ and $Y$.

If (and only if) the circuit current is fully emitter controlled, $L(X^{-1})$ is given by Eq. (2), and the related FN plot has slope $S(X^{-1})=\mathrm{d}L/\mathrm{d}(X^{-1})$. Thus, for fully emitter-controlled situations, Eq. (8) yields

$$\sigma_{YX}(X^{-1}) = -[\mathrm{d}\ln\{C\}/\mathrm{d}(X^{-1})]/S^* + v_F + X^{-1}\mathrm{d}v_F/\mathrm{d}(X^{-1}) + (v_F X^{-1}/S^*)\cdot \mathrm{d}(S^*)/\mathrm{d}(X^{-1}). \tag{9}$$

Customary practice is to then disregard the first and last terms on the right-hand-side of Eq. (9), which relate to effects such as field dependence in emission area, work function and emitter geometry. This constitutes the *basic approximation*.

In the basic approximation, the slope correction factor is denoted by $\sigma_B$ and given by

$$\sigma_B = v_F + X^{-1}\mathrm{d}v_F/\mathrm{d}(X^{-1}) = v_F - X\mathrm{d}v_F/\mathrm{d}X = v_F - F_C\mathrm{d}v_F/\mathrm{d}F_C. \tag{10}$$

The basic slope correction factor $\sigma_B$ applies to all types of FN plot. For any specific well-behaved barrier form, $\sigma_B$ can readily be found by numerical evaluation of barrier-strength integrals, as will be shown in a follow-up paper[12].

For the SN barrier used in orthodox emission, $v_F$ is given by $v_F$, where $v$ is the principal SN barrier function[10,13]. In this case,

$$\sigma_B = v_F - f_C dv_F/df_C \equiv s(f_C), \qquad (11)$$

where $s(f)$ is the slope correction function for the SN barrier, introduced by Burgess et al.[14] in 1953 as a function of the Nordheim parameter $y$ [$=\sqrt{f}$], but expressed here as a function of $f$ and tabulated as a function of $f$ in Ref. 2.

Formulas (9) to (11) assume that the emission process is CFE and the circuit current is fully emitter controlled. If this is not the case (as when series resistance causes saturation) then a slope correction factor is still formally defined by Eq. (8), but predictions of $S(X^{-1})$ and $\sigma_{YX}$ become much more complicated. In such situations a better approach may be to simulate circuit behaviour and compare it with experiment. Such situations are outside the scope of the present paper.

### B. Extraction of value of auxiliary constant $c_X$

The commonest use of FN plot analysis is to extract a value for the auxiliary constant $c_X$. Let the slope of a line fitted to the data points in a FN plot be $S^{\text{fit}}$. In the tangent approach, this fitted line is taken as parallel to the tangent to Eq. (2) at the value $X^{-1}=X_t^{-1}$. From Eqs. (5) and (8):

$$S^* = b\phi^{3/2}/c_X = -S(X_t^{-1})/\sigma_{YX}(X_t^{-1}) = S^{\text{fit}}/\sigma_t, \qquad (12)$$

where $\sigma_t \equiv \sigma_{YX}(X_t^{-1})$. Hence, if $\phi$ is known and $\sigma_t$ can be estimated, the extracted value is

$$c_X^{\text{extr}} = -\sigma_t b\phi^{3/2}/S^{\text{fit}}. \qquad (13)$$

At present, published $\sigma_t$-values exist only for orthodox emission. In this case, Eq. (13) becomes

$$c_X^{\text{extr}} = -s_t \cdot b\phi^{3/2} / S^{\text{fit}}. \tag{14}$$

It is shown elsewhere[8] that it is usually adequate to approximate $s_t$ as 0.95.

In LAFE literature, where usually $X$ is $F_M$ and $c_X$ is the field enhancement factor $\gamma_C$, customary practice is to use an elementary equation that omits $\sigma_t$. For orthodox LAFE emission situations, the resulting 5% error is rarely of significance. However, for non-orthodox situations, omission of $\sigma_t$ creates a defective equation that can yield[8] a spuriously large estimate for $\gamma_C$ (or, more generally, $c_X$).

## C. Extraction of *f*-values in orthodox emission situations

For orthodox emission, a formula related to those above, proved elsewhere[8], extracts the *f*-value ($f^{\text{extr}}$) related to a given experimental or simulated *X*-value ($X^{\text{expt}}$) as

$$f^{\text{extr}} = (s_t \eta / S^{\text{fit}}) X^{\text{expt}}. \tag{15}$$

This formula is used later in applying the test for lack of orthodoxy developed in Ref. 8, which is based on knowing an "apparently reasonable" range for extracted *f*-values.

## IV. THE 2012 INTERCEPT CORRECTION FACTOR

The need for an intercept correction factor arises when, working with Eq. (2), we wish to use the FN plot intercept to find a value for *C*. To understand the mathematical problem involved, initially treat *C* as constant. If $\nu_F$ in Eq. (2) were also constant, then finding *C* would be simple: $L(X^{-1})$ would be equal to $\ln\{C\}$ at $X^{-1}=0$, and, to find $C$, one would need only to extrapolate the FN plot to $X^{-1}=0$, as in undergraduate regression problems.

In reality, $\nu_F$ is not constant (except for physically unrealistic models based on the exact triangular barrier). In practice, $\nu_F$ goes to zero at some reference value $(X_R^{-1})$, and—if $C$ were constant—then $C$ would be given by $L(X_R^{-1})$. The extrapolated experimental FN plot intercepts the $L$-axis at a somewhat higher $L$-value, written here as $\ln\{I^{\text{fit}}\}$ ($I$ means "intercept-related parameter", not "current".)

The value of $L(X_R^{-1})$ cannot be determined by purely experimental means, because most emitters melt or explode for values of $X$ somewhat less than $X_R$. Hence, to use a "fitted line" approach, one needs to develop theory for interpreting $\ln\{I^{\text{fit}}\}$. This theory needs to be based on a barrier model, and numerical values of correction factors are different for different barrier models. The mathematical problem is further complicated because (for realistic barrier models) the theoretical function $L(X^{-1})$ in practice generates a curve rather than a straight line, and $C$ may depend on $X^{-1}$.

As noted above, the tangent method[2] of FN plot analysis assumes that the fitted line may be modelled by the tangent to the theoretical curve $L(X^{-1})$, taken at some specific (but initially unknown) value $X_t^{-1}$ at which the tangent would be parallel to the fitted line. Due to curvature in $L(X^{-1})$, the fitted straight line is expected to be a chord to $L(X^{-1})$ rather than a tangent. In principle, this can be dealt with by a "chord correction" (see Ref. 2 and below). However, typical values are not known for this correction, and it is not currently done as part of the tangent method, because the correction is thought to be small in comparison with other uncertainties. Hence, the parameter $I^{\text{tan}}$ below is currently put equal to $I^{\text{fit}}$.

<p style="text-align: center;">FIGURE 1 NEAR HERE</p>

For any particular values of the variables $\{X,Y\}$, the tangent to the theoretical curve $L(X^{-1})$, taken at $X^{-1}$, is defined to have slope $S(X^{-1})$ and intercept $\ln\{I^{\text{tan}}(X^{-1})\}$. The *2012 intercept correction factor* $\rho_{YX}(X^{-1})$ is then defined as

$$\rho_{YX}(X^{-1}) \equiv I^{\text{tan}}(X^{-1}) / C(X^{-1}) . \tag{16}$$

The mathematical problem is to find an expression for $\rho_{YX}(X^{-1})$. This is illustrated in FIG. 1 (where $C$

is written $C_{YX}$, as a reminder that the form of $C$ depends on the choice of $X$ and $Y$). Curve L is the theoretical curve $L(X^{-1})$, and line T is the tangent to L, taken at $X_t^{-1}$. Line T has slope $S(X_t^{-1}) = -\sigma_{YX}(X_t^{-1}) \cdot S*$ and intercept $\ln\{I^{\tan}(X_t^{-1})\}$. For clarity, the curvature in L is greatly exaggerated. Curve L terminates at point "R", where the barrier strength becomes zero.

This problem is most easily solved by a simple geometrical argument that finds $L(X_t^{-1})$ in two ways. (1) Directly from Eq. (2):

$$L(X_t^{-1}) = \ln\{C_{YX}(X_t^{-1})\} - v_F S*/X_t. \tag{17}$$

This evaluation is illustrated by line N in Fig. 1. (2) From tangent T:

$$L(X_t^{-1}) = \ln\{I^{\tan}(X_t^{-1})\} - \sigma_{YX}(X_t^{-1}) \cdot S*/X_t. \tag{18}$$

Equations (16) to (18) yield

$$\ln\{\rho_{YX}(X^{-1})\} = [\sigma_{YX}(X^{-1}) - v_F(X^{-1})](S*/X), \tag{19}$$

where the subscript "t" has been dropped because the result applies to any value $X^{-1} > X_R^{-1}$. Further, no need exists to continue to show the dependence on $X^{-1}$, so—using earlier relationships—this result can be written in the simpler forms

$$\ln \rho_{YX} = [\sigma_{YX} - v_F](b\phi^{3/2}/F_C) = [\sigma_{YX} - v_F]G_F^{ET}. \tag{20}$$

In the basic approximation defined earlier, result (20) reduces to

$$\ln \rho_B = (\sigma_B - v_F)(b\phi^{3/2}/F_C) = (\sigma_B - v_F)G_F^{ET}. \tag{21}$$

This result applies to any form of FN plot. For any particular choice of barrier model, $\rho_B$ is easily evaluated by numerical means, as will be shown in a follow-up paper[12].

For the SN barrier used in orthodox emission, $\rho_B$ is given by an intercept correction function that is a new type of SN barrier function and is denoted here by $r_{2012}$. This is given by

$$r_{2012} \equiv \exp[(s - v_F)(b\phi^{3/2}/F_C)] = \exp[-f_C(dv_F/df)(b\phi^{3/2}/F_C)], \qquad (22)$$

where the second step follows from the definition[10] of $s$. In scaled form, this becomes

$$r_{2012} = \exp[\eta u_F] \equiv \exp[\eta \cdot u(f_C)], \qquad (23)$$

where $u(f)$ is an SN-barrier function defined in Ref. 10 and tabulated in Ref. 2. Table 1 shows illustrative values of $r_{2012}$ that have been calculated exactly and rounded to integer values. Under normal working conditions $r_{2012}$ is typically around 100 to 150.

TABLE 1 NEAR HERE

In practice, a good working approximation[10] exists for $u(f)$:

$$u(f) \approx (5 - \ln f)/6. \qquad (24)$$

This gives results good enough for most technological purposes. [There is a typographical error in the more sophisticated formula given as Eq. (5.1) in Ref. 10: this should read: $u(l') \approx (1-q) - q\ln l'$.]

## V. ILLUSTRATION OF ORTHODOX EXTRACTION PROCEDURE

### A. Introduction

To apply the new approach to a FN plot, estimates are needed for the value $X_t$ at which the fitted line is parallel to the theoretical curve, and for $\sigma_{YX}(X_t)$ and $\rho_{YX}(X_t)$. There has been much theoretical work (e.g., Ref. 15) on predicting $Y(X)$ characteristics for various solid conductor shapes and emitter materials, but—except for orthodox emission situations—very little work on calculating slope and intercept correction factors. Hence, an orthodox situation is used to illustrate the new approach. In this case, one needs to find the scaled barrier field $f_t$ corresponding to $X_t$, and evaluate $s(f_t)$ and $r_{2012}(f_t)$.

Consider a LAFE analysis where $X$ is taken as $F_M$ and $Y$ as $J_M$. The full FN-type equation for $J_M(F_M)$ is[6]

$$J_M = \lambda_M^{SN} a\phi^{-1}\gamma_C^2 F_M^2 \exp[-v_F b\phi^{3/2}/\gamma_C F_M], \qquad (25)$$

where $a$ is the First FN Constant[9], and $\lambda_M^{SN}$ is the macroscopic pre-exponential correction factor for SN-barrier-based equations. Comparison with Eq. (1) shows that $c_X = \gamma_C$ and

$$C = \lambda_M^{SN} a\phi^{-1}\gamma_C^2. \qquad (26)$$

As noted earlier, the main ways of extracting a value for $\lambda_M^{SN}$ are the "spot-value method" used in Ref. 6, and the "fitted line" method described above. By re-using the Ref. 6 simulated data, the two methods can be compared.

FIGURE 2 NEAR HERE

Figure 2 here is a slightly modified version of Fig. 2 in Ref. 6. With $\phi$=4.50 eV, $\lambda_M^{SN}=10^{-9}$, $\gamma_C$=500, data points were calculated for $f$-values in the range $0.25 \leq f \leq 0.34$, at intervals of 0.005. This range corresponds (see Ref. 11) to the upper part of the voltage range used by Dyke and Trolan[16] in steady-current experiments on their emitter X89, in work to test the validity of FN theory. For $\phi$=4.50 eV, the reference field $F_R$= 14.06 V/nm; thus, this range of $f$ corresponds to $F_C$-values in {3.52 V/nm<$F_C$<4.78 V/nm}, to $F_M$-values in {7.03 V/µm<$F_M$< 9.56 V/µm}, and to ($1/F_M$)-values in {0.105 µm/V<$1/F_M$< 0.142 µm/V}. To assist checking, results (derived with a spreadsheet) are mostly given to 3 significant figures, but corresponding accuracy is not necessarily implied.

**B. Analysis of the slope**

For a "data analyst" who can assume orthodox emission and $\phi$=4.50 eV, but is otherwise unaware of simulation input parameters, the extraction process works as follows. By applying a ruler to Fig. 2 in Ref. 6, the slope $S^{\text{fit}}$ was found[6] as –123 Np V/µm (the neper is used as the unit of natural logarithmic difference). Using $\eta$= 4.637 and $s_t$= 0.950, and taking $X$ as $F_M$, Eq. (16) becomes

$$(1/f)^{\text{extr}} = (27.9 \text{ V/µm}) / F_M . \tag{27}$$

From Fig. 2 here, the data lie approximately in {0.1045 µm/V≤(1/$F_M$)≤0.142 µm/V}. Hence, the plot data correspond to values of $(1/f)^{\text{extr}}$ in {2.92≤$(1/f)^{\text{extr}}$≤3.96}, and to values of $f^{\text{extr}}$ in {0.252≤ $f^{\text{extr}}$≤0.343}. The mid-range value $[(1/f)^{\text{extr}}]_{\text{mid}}$ is 3.44, and its reciprocal (denoted here by $f_{t1}$) is 0.291.

As compared with the simulation inputs, the range endpoints have been extracted with an accuracy of about 1%. For the data-analyst, the extracted range {0.252≤$f^{\text{extr}}$≤0.343} means the emission is apparently orthodox, because this range lies totally within the range {0.15≤$f^{\text{extr}}$ ≤0.45} categorised as the "apparently reasonable" range of $f$-values, in the test described in Ref. 8.

A good working formula[10] for $s(f)$, accurate to better than 1%, is

$$s(f) = 1 - f/6 . \tag{28}$$

Hence, $s$-values applying to the plot lie is {0.943≤$s$≤0.958}, and $s(f_{t1})$=0.952. This compares well with the "typical value" 0.950 used in deriving Eq. (27), and is a necessary check of self-consistency.

It is possible to iterate the above analysis, using $s_t$=0.952 (rather than $s_t$=0.950) to evaluate the constant in Eq. (27), but the consistency gain is too small to be useful. The data analyst thus takes the fitting value $f_t$ as 0.291, and $s_t$ as 0.952.

Substitution into Eq. (15), with $c_X$ interpreted as $\gamma_C$, and using $b \approx$ 6831 eV$^{-3/2}$ V/µm, $\phi$= 4.50 eV,

$s_t$= 0.952, $S^{fit}$= –123 Np V/µm, yields the extracted value $\gamma_C^{extr}$≈ 504. This is satisfactorily close to the input value $\gamma_C$=500. Omission of the correction factor $s_t$, as in many LAFE papers, would yield the less accurate result $\gamma_C^{extr}$≈ 530.

**C. Analysis of the intercept**

From Fig. 2, extrapolating the fitted line by ruler to the left-hand boundary (1/$F_M$= 0.1 µm/V) yields $L$= –30.7 Np. Since $S^{fit}$= –123 Np V/µm, the fitted intercept ln{$I^{fit}$} = (–30.7+12.3)= –18.4 Np. If a chord correction, written as ln{$\rho^{chord}$}, were to be applied, then the experimental estimate of ln{$I^{tan}$} would be

$$\ln\{(I^{tan})^{expt}\} = \ln\{I^{fit}\} + \ln\{\rho^{chord}\} . \tag{29}$$

We disregard this correction here, and take $(I^{tan})^{expt}= I^{fit}$= exp[–18.4]= 1.02×10$^{-8}$ A V$^{-2}$.

From Eq. (24), the data analyst finds $u(f_t)= u(0.291)$≈ 1.039, and from Eq. (23) that $r_{2012}(f_t)$≈ 124. Definition (16) can now be applied in the inverse form

$$C^{extr} = (I^{tan})^{extr} / \rho_{YX}(X_t^{-1}) \approx I^{fit} / \rho_{YX}(X_t^{-1}) . \tag{30}$$

For orthodox emission, this reduces to

$$C^{extr} \approx I^{fit} / r_{2012}(f_t) . \tag{31}$$

Hence the analyst extracts the value $C^{extr}$= (1.02×10$^{-8}$)/124= 8.24×10$^{-11}$ A V$^{-2}$.

Inverting Eq. (26) yields

$$(\lambda_M^{SN})^{extr} = C^{extr} / [a\phi^{-1}(\gamma_C^{extr})^2]. \tag{32}$$

Using $a \approx 1.541$ µA eV V$^{-2}$, $\phi = 4.50$ eV, $\gamma_C^{extr} = 504$, the product $[a\phi^{-1}(\gamma_C^{extr})^2]$ is evaluated as $8.72 \times 10^{-2}$ A V$^{-2}$. Hence, the extracted value for $\lambda_M^{SN}$ is

$$(\lambda_M^{SN})^{extr} = C^{extr} / a\phi^{-1}(\gamma_C^{extr})^2 = (8.24 \times 10^{-11} / 8.72 \times 10^{-2}) = 9.45 \times 10^{-10}. \tag{33}$$

This compares well with the $\lambda_M^{SN}$ input value of $10^{-9}$, and shows that the new approach introduced in this article works, at least for orthodox emission and for the data set used here and in Ref. 6.

### D. Comments

Considering that a ruler, rather than a regression calculation, has been used to fit a line to the data points, the closeness of the agreement between input and extracted values is perhaps better than one might reasonably expect, and may involve an element of luck (possibly helped by the regularity of simulated data points).

For an SN barrier the curvature of $L(X^{-1})$ is greater at the left-hand-side (low-$f^{-1}$ side) of a FN plot (see Ref. 2). This means that, strictly, the fitting position should be taken somewhat to the left of the mid-range value of $X^{-1}$. An equivalent effect was shown in Ref. 17 (using the old approach), when extracting notional emission area. In the methodology here, the advantages of a marginally more accurate extraction procedure are outweighed by the merits of having an exactly defined procedure, namely that the fitting point should be taken at the experimental mid-range value of $X^{-1}$. However, when extraction theory is extended to non-orthodox emission, this issue will need to re-examining.

This simulation is presented as "illustration of procedure" and as "proof of concept". Although the extraction procedure should work adequately for other orthodox emission situations, the results here do not imply that it would work with equally good precision in all orthodox emission situations.

Operationally, all the calculations are readily programmed on a spreadsheet; it is then only a matter of entering data, including the results of fitting a line to the FN plot being analyzed.

## VI. DISCUSSION

### A. Comparison of new and old types of intercept correction factor

A change from earlier treatments is the use of a new physical quantity (the "2012 intercept correction factor"). The new quantity does not include any factors that form part of the pre-exponential of a FN-type equation, whereas the old form of definition did. For example, for orthodox emission the new intercept correction function is $r_{2012} = \exp[\eta u_F]$, whereas the old intercept correction function (denoted here by $r_{1999}$) defined in Ref. 3 and discussed in Ref. 10 was $r_{1999} = t_F^{-2} \exp[\eta u_F]$, where $t_F^{-2}$ is the local pre-exponential correction factor in the Murphy-Good FN-type equation[18]. Use of a different SN-barrier-based FN-type equation, with a slightly different pre-exponential factor, would have yielded a slightly different mathematical function. The new definition allows proper discrimination between pre-exponential correction factors related to the forms of FN-type equations and the correction related to the peculiar mathematics of FN plots of FN-type equations.

When the basic approximation above is reasonably valid, the new definition allows the same factors $\sigma_B$ and $\rho_B$ to be used for all types of FN plot, whereas the old approach would, in some cases, have needed different factors for different plot types. A further merit is that, when the basic approximation is valid, the new intercept correction factor can be calculated accurately, because it does not include the uncertainties associated with the pre-exponentials in realistic FN-type equations.

### B. Related methods

The main high-level alternative to the tangent method is the spot-value method used in Ref. 6,

which for the same data set yielded ($\lambda_M^{SN}$)$^{extr}$= 7.7×10$^{-10}$. One might expect a fitted-line result to be more accurate, but the difference is not significant in technological contexts (particularly at present, when many papers erroneously omit[6] a pre-exponential correction factor from their equations). In fact, having both methods is useful, as they serve as a mutual check.

For orthodox emission, other (older) methods of FN plot analysis also exist (see Ref. 2). These are linearization methods that first involve creating, for $L(X^{-1})$, a linear formula that represents either the equation itself or the tangent to it or (with the Spindt approximation[19]) a chord to it, and then applying line-fitting techniques. In effect, these methods create approximations to the intercept correction function $r_{2012}$, valid for some specific value of $f_t$. The tangent method should usually be slightly more accurate than these older methods (it would certainly be more accurate if a chord correction were made), and—now that simple good working formulas are known for $s(f)$ and $u(f)$—is no more difficult to use than the older methods.

However, the tangent method's main merit is its greater generality. It can deal with variations of $f_t$ as between different experimental data sets, and is more easily adapted to deal with barrier forms more general than the SN barrier.

In older literature, the SN-barrier functions are often expressed as functions of the Nordheim parameter $y$ [=$\sqrt{f}$] rather than scaled barrier field $f$. There are good mathematical and physical reasons why using $f$ is preferable, not least the greater transparency in the physical meaning of $f$ and the linearity of the relationship between $f$ and barrier field $F$. We strongly recommend that use of $y$ as a mathematical argument in CFE theory should be discontinued.

**C. Non-linear curve fitting**

In orthodox emission, FN plots are expected to be very slightly curved[2]. This raises the possibility of using non-linear curve-fitting procedures partly analogous to those used by Edgcombe and de Jonge[20] for spherical emitters. Our view is that for orthodox emission the curvature is so slight that it would usually be masked by data noise. Even if this problem could be overcome, there is a second difficulty. This arises because, for theoretical plots based on the SN barrier, the curvature in $L(X^{-1})$, although absolutely small, varies strongly with $X^{-1}$ (see Ref. 2). Consequently, the interpretation of any fitted curvature coefficient would be problematic unless one knew what part of the theoretical curve (what $f$–value range) the experimental plot corresponded to. It would probably be useful to first carry out linear fitting as described above, to estimate the $f$-value range involved.

Equivalent interpretational problems presumably exist for significantly curved emitters, to which orthodox emission theory does not apply. As FN plot analysis theory is extended to small-tip-radius emitters, beyond the approach of Edgcombe and de Jonge, problems of this kind may need exploring in much greater depth.

### D. Other plotting methods

FN plots are the commonest way of presenting CFE data graphically. FN plots are used because Stern et al. found[1] in 1929 (when re-plotting a current-field curve published by Millikan and Eyring[21] in 1926) that a FN plot gave a slightly better straight line than did the original 1928 method—due to Millikan and Lauritsen[22] (ML)—of plotting $\log_{10}\{i\}$ vs $V^{-1}$.

Modern theory[23] suggests that neither form is exactly correct. The best linear form is probably $[\ln\{i/V^\kappa\}$ vs $V^{-1}]$, where $\kappa$ is an unknown number, between $-1$ and $3$ and not necessarily integral, and possibly varying slightly with $V^{-1}$. For analysis corresponding to $V^\kappa$ in the pre-exponential, one would replace Eq. (1) by

$$Y_\kappa = C_\kappa X^\kappa \exp[-v_F S^*/X], \tag{34}$$

and define a new plotting form by

$$L_\kappa(X^{-1}) \equiv \ln\{Y_\kappa/X^\kappa\} = \ln\{C_\kappa\} - v_F S^* \cdot X^{-1}. \tag{35}$$

The remainder of the theory would resemble that developed above. In particular, the basic approximation would apply to these revised plotting methods.

**E. Future development**

The simulation above relates to orthodox emission and FN plots, because an immediate aim was to fill a gap in orthodox analysis theory. However, the assumption of orthodox emission is physically restrictive: many real emission situations are slightly or substantially non-orthodox. Because many equations in Sections III and IV are more general than orthodox emission requires, an obvious research path is to explore the effects of not making some of the approximations made above.

The great complexity of the issues involved (which is hidden in this article) means this needs to be done in stages. An obvious first stage is to stay within the basic approximation, but explore how different emitter shapes and barrier forms affect slope and intercept correction factors. Some completed initial studies on a spherical emitter, which rely on the theory here, will be submitted as a follow-up paper[12].

Beyond this, there is a need to explore effects related to the neglected terms in Eq. (9), which are both composite terms, and beyond this again to explore effects arising when the circuit current is not fully emitter controlled and analysis of circuit behaviour is required.

**F. Summary and Conclusions**

In relation to the tangent method of FN plot analysis, this paper has achieved the following. (a) It has introduced a new, more useful, type of intercept correction factor ($\rho_{YX}$), defined by Eq. (16). This enables the equation for the tangent to the theoretical curve in a "universal" FN plot to be written

$$\ln\{Y/X^2\} = \ln\{\rho_{YX}C\} - \sigma_{YX}S^* \cdot X^{-1} . \tag{36}$$

This has a clear relationship with Eq. (2), the equation for the theoretical curve itself. (b) For fully emitter-controlled situations, it has provided general formulas—Eqs (9) and (20)—for the slope correction factor and the new type of intercept correction factor. (c) By making a series of well-defined simplifying assumptions and approximations, it has shown how to reduce these general formulas to simple explicit formulas applying to orthodox emission and orthodox data analysis. (d) For orthodox emission, it has defined a new intercept correction function ($r_{2012}$), and has given an easily evaluated "scaled" formula for $r_{2012}$. (e) It has shown how to apply the improved orthodox data-analysis method to simulated orthodox emission data, and has shown that simulation input-parameter values (including the pre-exponential correction factor $\lambda_M^{SN}$) are recovered with good accuracy.

The tangent approach to FN-plot interpretation is superior to other orthodox methods, both because it is more flexible (and in principle slightly more accurate) for orthodox emission, and because it can be applied to many non-orthodox situations (including some smaller-radius emitters). We suggest that having a standard method of FN plot analysis would be helpful, and that for orthodox data analysis this improved tangent method should be used in preference to older methods.

Finally, we make the point that the simulation data used here and in Ref. 6 were prepared by assuming orthodox emission. Section V shows that orthodox emission situations can be successfully and self-consistently analysed by using the improved orthodox data-analysis method derived above. However, it remains true that applying orthodox data-analysis methods to non-orthodox situations can generate spurious results. Hence, the test for lack of orthodoxy developed in Ref. 8 needs to become an essential part of orthodox data-analysis procedure.

It is likely[8] that this test will show that many modern emission situations are non-orthodox. The general theory concerning $\sigma_{YX}$ and $\rho_{YX}$, in Sections II to IV above, is intended to enable research into data-analysis theory for non-orthodox situations. For such situations, it is believed that the orthodox procedure outlined in Section V could often be used as a paradigm, but with relevant slope and intercept correction factors calculated without making all the assumptions and approximations used to derive orthodox analysis theory.

Good data-analysis theory for non-orthodox emission situations is urgently needed. It is hoped that the more general aspects of this article will enable and open up its development.

**ACKNOWLEDGMENTS**

Andreas Fischer thanks the Alexander von Humboldt foundation for a Feodor Lynen fellowship and Mu'tah University for hospitality.

**Table 1**

TABLE I. Values of the 2012 intercept correction function ($r_{2012}$) for the Schottky-Nordheim barrier, as a function of local work function $\phi$ and scaled barrier field $f$, for practical ranges of both parameters. Values of $\eta$ as given by Eq. (7) are also shown.

| $\phi$ (eV) = | 3.00 | 3.50 | 4.00 | 4.50 | 5.00 | 5.50 |
|---|---|---|---|---|---|---|
| $\eta$ = | 5.679 | 5.258 | 4.918 | 4.637 | 4.399 | 4.194 |
| $f$ | | | | | | |
| 0.15 | 690 | 425 | 287 | 208 | 158 | 125 |
| 0.20 | 518 | 326 | 224 | 164 | 127 | 101 |
| 0.25 | 416 | 266 | 185 | 137 | 107 | 86 |
| 0.30 | 348 | 225 | 159 | 119 | 93 | 75 |
| 0.35 | 300 | 196 | 140 | 105 | 83 | 68 |
| 0.40 | 264 | 174 | 125 | 95 | 75 | 61 |
| 0.45 | 236 | 157 | 113 | 87 | 69 | 57 |

**Figure Captions**

FIG. 1. Schematic diagram illustrating the mathematics of the intercept correction factor $\rho_{YX}$, as defined by Eq. (16) (but with the symbol $C_{YX}$ used instead of $C$). Curve L represents any non-elementary FN-type equation, plotted in FN coordinates of type [$\ln\{Y/X^2\}$ vs $X^{-1}$)], and has the functional form $L(X^{-1})$. The curvature of L is greatly exaggerated. Curve L cuts off at point "R", at which the field-reduced barrier height becomes zero. Line T is the tangent to curve L, taken at the "fitting value" $X_t$. Line N shows how $L(X_t^{-1})$ is determined from curve L, when the pre-exponential of the FN-type equation has a value written in the form $C_{YX}(X_t^{-1})X_t^2$. Line T intersects the vertical axis at value $\ln\{I^{\tan}(X_t^{-1})\}$, and provides a second method of determining $L(X_t^{-1})$. The parameters $C_{YX}$, $I^{\tan}$, $\sigma_{YX}$ and $\rho_{YX}$ are all evaluated for $X=X_t$.

FIG. 2. Simulated $J_M$-$F_M$-type FN plot based on Eq. (25), with input-parameter values as given in the text.